%% file: ms.tex
\documentclass[a4paper,12pt]{article}

\usepackage[latin2]{inputenc}
\usepackage[dvips]{graphicx}
\usepackage{amsmath}
\usepackage{amssymb}
\usepackage[mathcal]{eucal}
\usepackage[small,bf]{caption2}
\usepackage{fancyhdr}
\usepackage{natbib}
\bibpunct[, ]{(}{)}{,}{a}{}{,}
\usepackage{amsmath}

\newcommand{\refp}[1]{Figure \ref{#1}} 
\newcommand{\refe}{\eqref} 
\newcommand{\beq}{\begin{equation}}
\newcommand{\eq}{\end{equation}}
\newcommand{\cn}{\textrm{cn}}
\newcommand{\sn}{\textrm{sn}}
\newcommand{\dn}{\textrm{dn}}
\newcommand{\Th}{\tanh}
\newcommand{\Cth}{\coth}
\newcommand{\Arth}{\textrm{Artanh}}
\newcommand{\ArCth}{\textrm{Arcoth}}
\newcommand{\Ch}{\cosh}

\newcommand{\pp}{\mathcal{P}}
\newcommand{\pfi}{p_{\varphi}}
\newcommand{\ptt}{p_t}
\newcommand{\pth}{p_{\theta}}
\newcommand{\esp}{\hspace{1cm}}

\title{\Large \bf Optics in the Schwarzschild space-time}

\author{Andrej \v{C}ade\v{z}\footnote{e-mail: andrej.cadez@fmf.uni-lj.si} \ 
and Uro\v{s} Kosti\'{c}}

\date{}

\begin{document}
\maketitle
\begin{center}
\emph{
Department of Physics, Faculty of Mathematics and Physics, University of
Ljubljana\\
Jadranska 19, 1000 Ljubljana, Slovenia
}
\end{center}
\vspace{0.5cm}
\begin{abstract}
Realistic modelling of radiation transfer in and from variable
accretion disks around black holes requires the solution of the
problem: find the constants of motion and equation of motion of a
light-like geodesic connecting two arbitrary points in space. Here we
give the complete solution of this problem in the Schwarzschild space-time.
\end{abstract}
\vspace{0.5cm}
\section{Introduction}
	\input{ms1.tex}

\section{Connecting two points with a light-like geodesic}
	\input{ms2.tex}
\section{Calculation of the time of flight}
	\input{ms3.tex}
\section{Conclusions}
	\input{ms4.tex}
\bibliographystyle{plainnat}
\bibliography{biblio}
\end{document}

%% file: ms1.tex
The first light detected from a relativistic region about a black hole
was discovered by the ASCA satellite \citep{tanaka,yaqoob}. The now
accepted theoretical model describing the broad X-ray emission lines
is that of an accretion disk around either a Kerr or a Schwarzschild
black hole
\citep{1995MNRAS.277..901R,1997ApJ...488..164T,1997ApJ...477..602N,fanton,bromley,dabrowski,cadez2,martocchia}. 
This discovery increased the interest for phenomena occurring in the
vicinity of black holes. We now know that other interesting high
energy phenomena, such as X-ray flares
\citep{2001Natur.413...45B,2003ApJ...584..751G} or quasi-periodic oscillations
\citep{2001ApJ...552L..49S,2002ApJ...570L..69M} are occuring in the
environment of black holes. To further the understanding of such
phenomena from the theoretical point of view, it is important to
develop tools to model the phenomena themselves as well as to model
radiation transfer in and from these strongly curved regions of space-time.

Most current accretion disk models are very simple from the radiation
transfer point of view. They consider disks as geometrically thin and
optically thick. Light that reaches a far observer, comes directly
(without beeing scattered) from a very definite point on the disk. Therefore, line
profiles can be calculated  by aiming
light-like goedesic  from a point on the disk to the observer or vice
versa, aiming from the observer to points on the disk. In such
ray-tracing procedures geodesic equations are usually solved by direct
numerical integration
\citep{laor,2001ApJ...547..649P,1997ApJ...488..109R}. However, when
modelling transient phenomena produced by small debris around black
holes, or by other transient phenomena in the disk (moving hot spots,
varying external illumination, waves
), it
becomes necessary to solve a more difficult radiation transfer
problem, the problem of following a single photon through its more
than one scattering and/or more than one possible path from the source
to the eye of the observer
\citep{2005astro.ph..2048S,2004ApJ...606.1098S}. The problem is further
complicated by noting that photon arrival times from the same initial
source to the observer may (and often will) be markedly different for
photons reaching the observer along different possible paths. It is
obvious, that 
a multiple scattering path cannot be effectively constructed
by aiming
geodesics between succesive scattering points 
as the number of 
succesive iterations required would soon blow up.
The solution to the
radiation transfer problem 
thus requires 
one to be able to 
follow a
light ray 
from one to the other scattering point along its path.
Thus, one
would like to find the quickest way to determine all constants of
motion of a geodesic that connects the given initial and final point.
In this article we give analytic expressions which completely solve this
problem in the Schwarzschild space-time. We also turn this analytic
tool into a numerical code and demonstrate that it is much faster,
more accurate and more transparent than aiming and integrating geodesic
equations. This tool thus 
opens the possibility to solve complex radiation
transfer problems in curved space-time using similar Monte Carlo
techniques that are used in solving radiation problems in flat space-time. 

Our work starts with the results of 
\citet{chandra} and
\citet{rauch}, who expressed the solutions to
geodesic equations in terms of elliptic integrals. 
By inverting their expressions into Jacobi elliptic functions
\citep{cadez2}, we obtain simple solutions for all the three types of
orbit equations that occur in Schwarzscild space-time. These solutions
no longer contain branch ambiguities. Since these orbits are
essentialy planar, their equations depend only on two nontrivial
constants of motion: the angular momentum and the longitude of the
periastron. Expressing the orbit equation at the initial and final
point on the geodesic, one obtains two nonlinear equations for the two
nontrivial constants of motion. However, since the longitude of the
periastron occurs only linearly as the argument of elliptic functions,
it is possible to use the elliptic functions addition theorem to
elliminate the longitude of the periastron and obtain a single
nonlinear equation for the angular momentum as function of initial and final coordinates.  
Here we derive these equations for all the three types of orbits and
discuss their properties and solutions. We also write down all the
other constants of motion in terms of final and initial point
coordinates and give analytic expressions for travel times.

%% file: ms2.tex
In the Schwarzschild space-time it is customary to introduce
Schwarzscild coordinates $t$, $r$, $\theta$ and $\varphi$. In these coordinates geodesics are governed by the Hamiltonian:
\beq
 H~=~\frac{1}{2}
   \Biggl[
     -\frac{1}{1-\frac{2M}{r}}\ \ptt^2 + \Bigl(1-\frac{2M}{r}\Bigr )p_r^2 +
      \frac{1}{r^2} \Bigl(\pth^2 + \frac{1}{\sin^2\theta} \ \pfi^2\Bigr )
   \Biggr ]~~, \label{hamiltonian}   
\eq
which admits 8 constants of motion: the value of the Hamiltonian ($H$)
and the value of the Lagrangian ($L$) (the ratio of the two defining
the relation between time and proper time), the energy $E=\ptt$, the
angular momentum ($\vec{l}$), the longitude of the periastron
($\omega$) and the time of periastron passage.

The angular momentum is expressed as $\vec{l}=l\cdot\hat{n}$, where
$\hat{n}$ is the unit vector along $\vec{l}$, which is defined by the
inclination of the orbit ($\varepsilon$) and the longitude of the
ascending node ($\Omega$).

Consider a light-like geodesic connecting the initial point $\pp_i$
and the final point $\pp_f$. Five constants $H=0$, $L=0$, $E$,
$\Omega$ and $\varepsilon$ are determined readily. So one can use the
true anomaly $\lambda$ as a parameter along the
geodesic. From the initial to the final point $\lambda$ increases
by:
\beq
  \Delta \lambda_f = \delta \lambda + 2\pi k\ ,
\eq
where 
\beq
\delta \lambda = \arccos \big[ \cos\theta_i \cos\theta_f + \sin\theta_i
    \sin\theta_f \cos(\varphi_f - \varphi_i)\big] \ ,
\label{lambda}
\eq
and the winding number $k={...-1,0,1...}$ tells how many times a
geodesic winds around the black hole.\footnote{Note, if $\Delta \lambda_f < 0$, replace $\Delta
\lambda_f \rightarrow -\Delta \lambda_f$ and $\hat{n} \rightarrow -\hat{n}$.}

\begin{figure}
\begin{center}
\includegraphics[width=6cm]{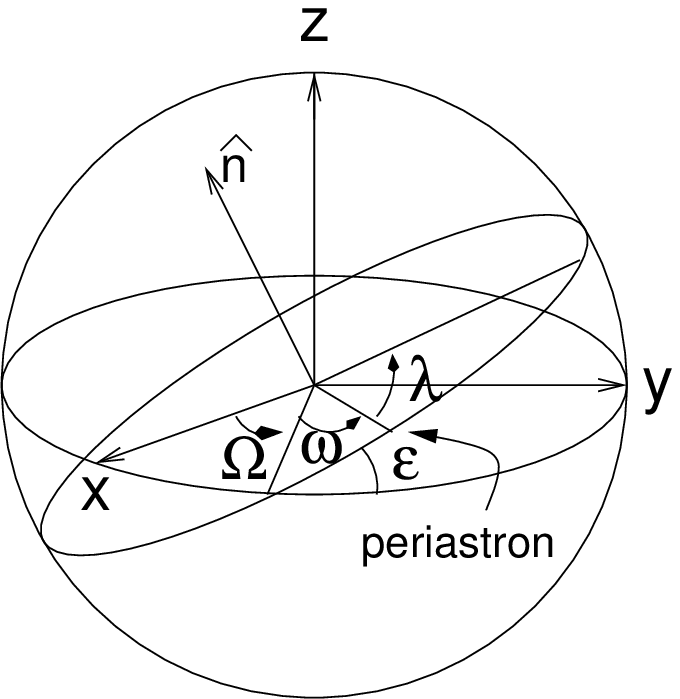}
\hspace{2cm}
\includegraphics[width=6cm]{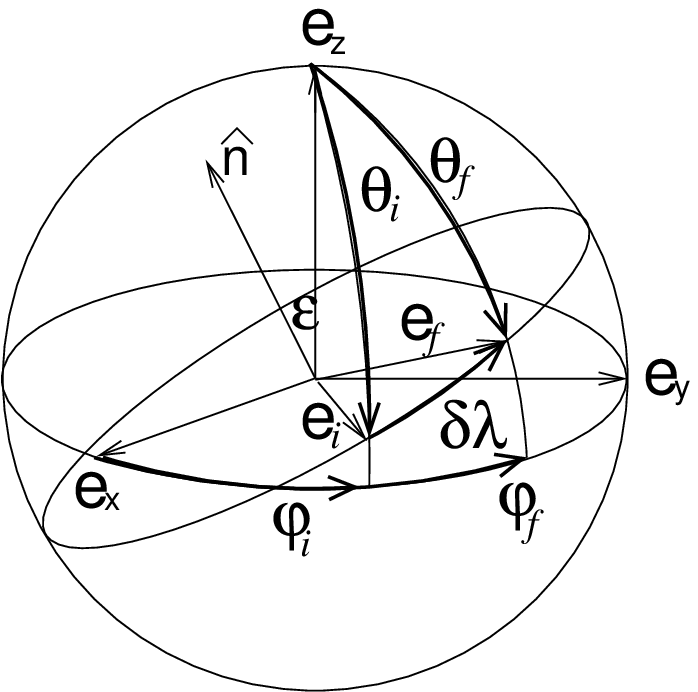}
\end{center}
\caption{Left: The orbital plane in equatorial coordinates: $\hat{n}$ unit normal, $\varepsilon$
inclination, $\Omega$ longitude of the ascending node, $\omega$
longitude of the periastron and $\lambda$ the true anomaly. Right: The
initial $\pp_i=(\theta_i,\varphi_i)$ and the final
$\pp_f=(\theta_f,\varphi_f)$ point.}
\label{koti}
\end{figure}

\begin{figure}
\begin{center}
\includegraphics[width=\textwidth]{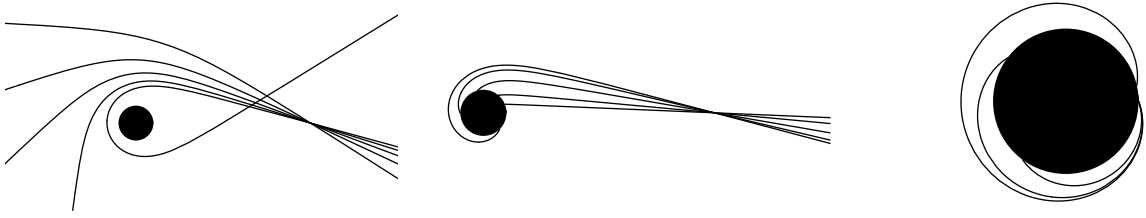}
\end{center}
\caption{Left: Some orbits of type \emph{A} with parameters $r_i = 20
M$ and $a_{crit}-a \in \lbrace 0.2, 0.15, 0.1, 0.05,
0.005\rbrace$. Middle: Orbits of type \emph{B} with parameters $r_i = 20
M$ and $a-a_{crit} \in \lbrace 2, 0.7, 0.2, 0.05,
0.005\rbrace$. Right: Orbits of type \emph{C} with parameters $r_i = 2.00001
M$ and $a_{crit}-a \in \lbrace 0.01, 0.005, 0.001 \rbrace$. The radius
of the black circle is the Schwarzschild radius.}
\label{orbite}
\end{figure}

\begin{figure}
\begin{center}
\fbox{\includegraphics[width=5.5cm]{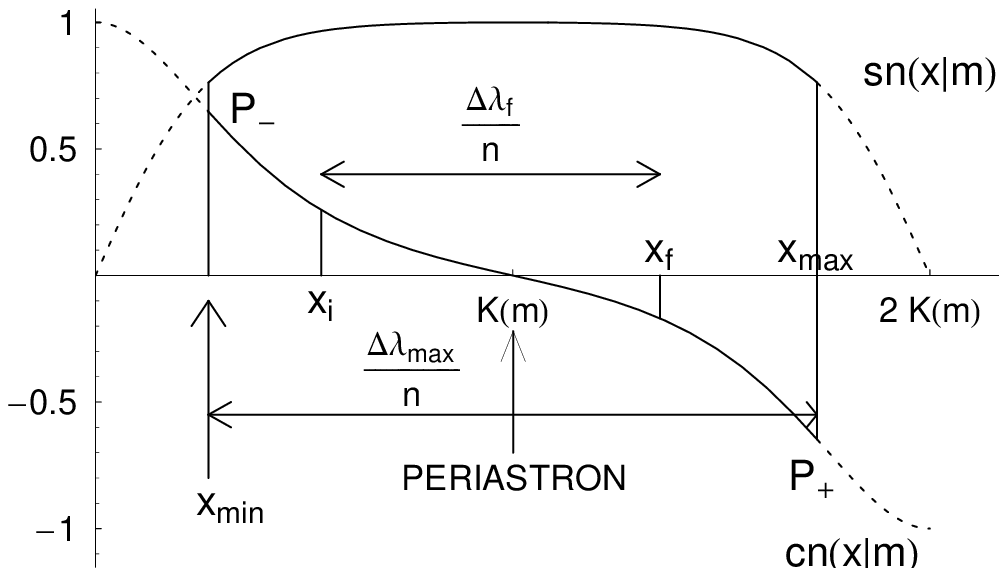}}
\fbox{\includegraphics[width=5.5cm]{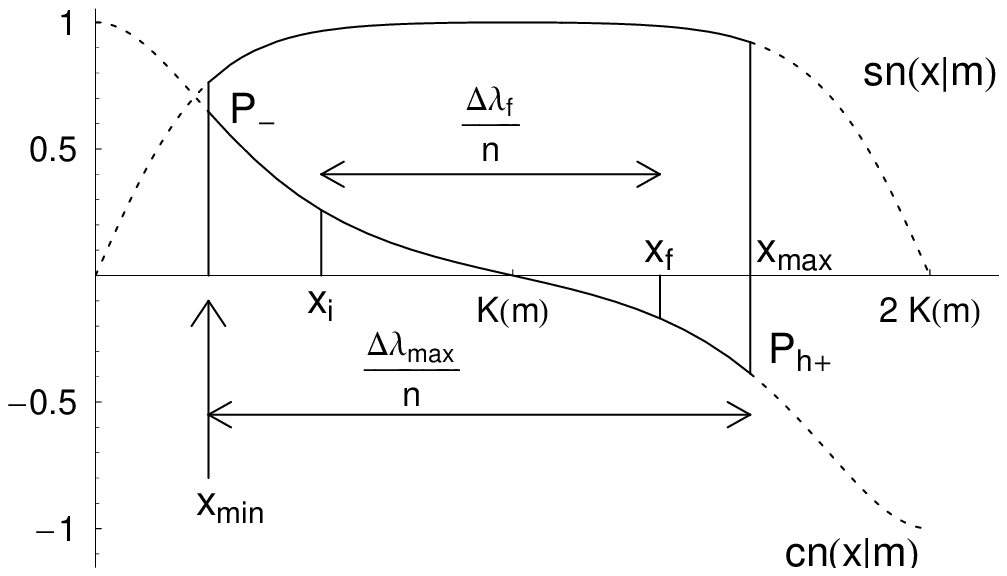}}
\fbox{\includegraphics[width=5.5cm]{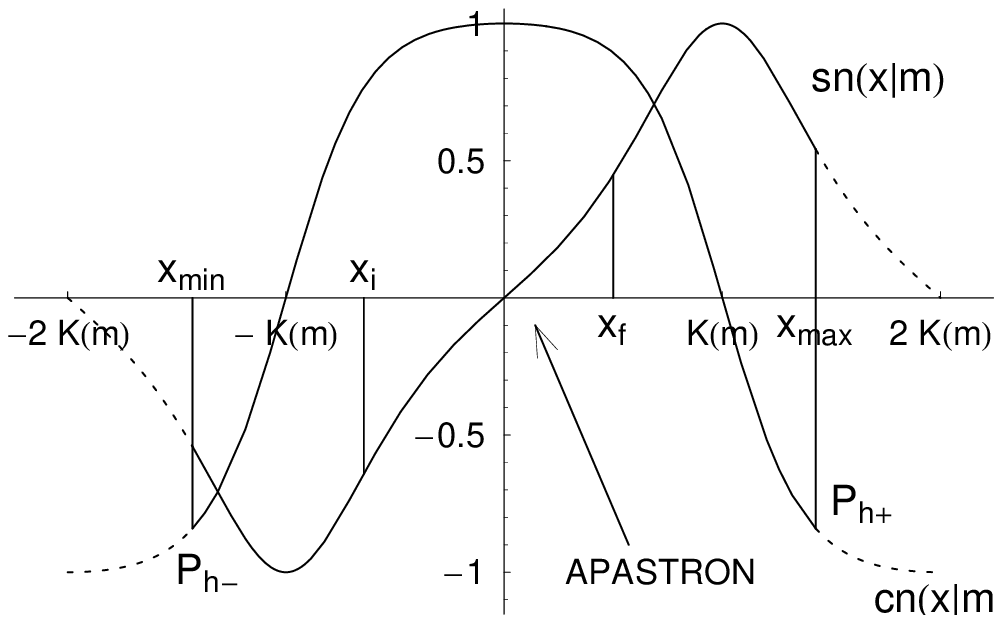}}
\end{center}
\caption{Functions $\cn(x|m)$ and $\sn(x|m)$ along orbits: left type
\emph{A}, middle type \emph{B}, right type \emph{C}. $x_{min}$ and
$x_{max}$ are at the endpoints of the orbit and are designated by
$\pp_-$ and $\pp_+$ if they are at spatial infinity, and $\pp_{h-}$
and $\pp_{h+}$ if they are at the horizon of the black hole. Interval
definitions for the endpoints follow after equations \refe{orbitaA},
\refe{orbitaB} and \refe{orbitaC}. $x_i=z$ and $x_f=v$ are at the
initial and the final point of the orbit.}
\label{jacobi}
\end{figure}

The two constants $\Omega$ and $\varepsilon$ are obtained from
angular coordinates of the initial and final point (see \refp{koti})
with the help of basic spherical trigonometry:
\begin{align}
  \cos \varepsilon & =\frac{
       \sin\theta_i \sin\theta_f \sin(\varphi_f - \varphi_i)
       }
       {
       \sin\delta\lambda}
       \\
  \tan\frac{\Omega}{2} & =\frac{
				\sin \theta_i \cos \theta_f \sin \varphi_i -
				\cos \theta_i \sin \theta_f \sin \varphi_f
				}
  				{
				\sin \varepsilon \sin \delta \lambda +
				\sin \theta_i \cos \theta_f \cos \varphi_i -
				\cos \theta_i \sin \theta_f \cos \varphi_f
				} \ .
\end{align}
Since the variable $\lambda$ is conjugate to the orbital angular
momentum, it obeys the Poisson bracket relation $[\lambda,l]=1$, so
that the angles $\theta$ and $\varphi$ are expressed with $\lambda$ as
follows \citep[eqs.~(6 -- 16)]{cadez3}:
\begin{align}
  \cos\theta & = -\sin\varepsilon \sin (\lambda+\omega)\\
  \tan \frac{\varphi -\Omega}{2} & = \frac{\cos\varepsilon \sin(\lambda + \omega)}
  					  {\sin\theta + \cos(\lambda + \omega)}
\end{align}
and\footnote{Note, that $\sin\theta=+\sqrt{1-\cos^2\theta}$.} the differential equation for orbits of light-like geodesics becomes:
\beq
  \frac{d u}{d\lambda} = \pm\sqrt{a^2-u^2(1-u)} \ .
\label{orbita}
\eq
Here $u=2 M/r$ and $a=2ME/l$. The solutions to this equation, called orbit equations, depend
only on the parameter $a$ and are of three types (\refp{orbite}):
\begin{itemize}
\item[-]\textbf{type \emph{A}}: scattering
orbits with both endpoints at infinity; their angular
momentum parameter is on the interval $0<a<\frac{2}{3 \sqrt
3}$. Scattering orbits can never extend below $r=3M$. 
\item[-]\textbf{type \emph{B}}: plunging orbits with one end at infinity and the other
behind the horizon; $a>\frac{2}{3 \sqrt 3}$, 
\item[-]\textbf{type \emph{C}}: near orbits with both ends behind the horizon of the black
hole; their angular momentum parameter is on the interval
$0<a<\frac{2}{3 \sqrt 3}$. Near orbits can never reach beyond $r=3M$. 
\end{itemize}
For completeness, we list the solutions of the orbit equation \refe{orbita} in terms of a few auxiliary parameters.
%
%
\\
\\
{\bf Type \emph{A}:}
Introduce $a= \frac{2}{3\sqrt{3}} \sin \frac{\psi}{2}$ ($0<\psi <\pi
$) and the following auxiliary parameters, functions of $\psi$ only:
%
\begin{subequations}
\begin{align}
    u_1 & = \frac{1}{3}\Bigl( 1+2\cos\frac{\psi}{3}  \Bigr) \label{u1}\\
    u_2 & = \frac{1}{3}\Bigl( 1+2\cos\frac{\psi-2\pi}{3}  \Bigr)\label{u2}\\
    u_3 & = \frac{1}{3}\Bigl( 1+2\cos\frac{\psi+2\pi}{3}  \Bigr) \label{u3}\\
      m & = \frac{u_2-u_3}{u_1-u_3}\\
      n & = \frac{2}{\sqrt{u_1-u_3}}\label{nA}\\
 \chi_i &{\rm ~~such~ that~}\  u_i=u_2-(u_2-u_3)\cos^2\chi_i\ ,
\label{defHiA}
\end{align}
\end{subequations}
%
where $u_i=2M/r_i$ and similarly $u_f=2M/r_f$. In terms of these, the
type \textbf{{\emph {A}}} differential equation for the  orbit becomes (\refp{orbite}, left):
\beq
   u = u_2-(u_2-u_3)\cn^2
        \bigl(
	      F(\chi_i|m) + \frac{\Delta\lambda}{n}|m
	\bigr)\ , \label{orbitaA}
\eq
with $\Delta\lambda = \lambda - \lambda_i$ and $\Delta\lambda /n \in \bigl \lbrace x_{min}-x_i~,~(2
K(m)-x_{min})-x_i\bigr\rbrace$, where
$x_{min}=F(\arccos\sqrt{u_2/(u_2-u_3)}\vert m)$ and
$x_{i}=F(\arccos\sqrt{(u_2-u_i)/(u_2-u_3)}\vert m)$, see \refp{jacobi}
left. Here $F$ and $K$ are the incomplete and complete elliptic
integrals of the first kind \citep{mathematica}.
%
%
\\
\\
{\bf Critical \emph{A}:}
A special limiting case of solution \refe{orbitaA} for $\psi = \pi$
($a=\frac{2}{3 \sqrt 3}$) and $u_i<2/3$ ($r_i>3M$) is:
\beq
  u = -\frac{1}{3} +\Th^2
     \bigl(
           \Arth\sqrt{u_i+1/3}~+\frac{\Delta\lambda}{2}
     \bigr) \ ,
\label{orbitaK}
\eq
with $\Delta\lambda = \lambda - \lambda_i$ and $\Delta \lambda ~\in
~\bigl \lbrace 2 \Arth\sqrt{1/3}-2 \Arth\sqrt{u_i+1/3}~,~\infty \bigr\rbrace$.
%
%
\\
\\
{\bf Type \emph{B}:}
Introduce $a= \frac{2}{3\sqrt{3}}\cosh \mu $  ($0<\mu <\infty$)
and the auxiliary 
parameters, functions of $\mu$ only:
%
\begin{subequations}
\begin{align}
    u_1 & = \frac{1}{3}\Bigl( 1-2\Ch\frac{2\mu}{3}  \Bigr) \label{u1B}\\
      m & = \frac{1}{2}\left( 1- \frac{3u_1-1}{2\sqrt{u_1(3u_1-2)}} \right)\\
      n & = \bigl( u_1(3u_1-2)\bigr)^{-\frac{1}{4}} \label{nB}\\
 \chi_i & {\rm ~~such~that~}\  u_i=u_1+\frac{1}{n^2}\hspace{1pt}\frac{1-\cos \chi_i}{1+\cos \chi_i} \label{hib} \ .
\end{align}
\end{subequations}
%
The type \textbf{{\emph {B}}} differenial equation for the orbit is (\refp{orbite}, centre):
\beq
u = u_1 + \frac{1}{n^2}\hspace{1pt}
        \frac{ 
	      1-\cn\bigl( F(\chi_i|m) + \frac{\Delta\lambda}{n}|m
	      \bigr) 
	     }{
	      1+\cn\bigl( F(\chi_i|m) + \frac{\Delta\lambda}{n}|m
	      \bigr) 
	      } \label{orbitaB} \ ,
\eq
with $\Delta\lambda = \lambda - \lambda_i$ and $\Delta
\lambda/n~\in~\bigl \lbrace F(\chi_{BH}\vert m)-F(\chi_i\vert
m)~,~F(\chi_{\infty}\vert m)-F(\chi_i\vert m) \bigr\rbrace$, where
$\chi_{BH}=\arccos\frac{1-n^2(1-u_1)}{1+n^2(1-u_1)}$ and
$\chi_{\infty}=\arccos\frac{1+n^2u_1}{1-n^2u_1}$.
%
%
\\
\\
{\bf Type \emph{C}:}
Introduce $a= \frac{2}{3\sqrt{3}} \sin \frac{\psi}{2}$ ($0<\psi <\pi
$) and the auxiliary parameters, functions of $\psi$ only:
%
\begin{subequations}
\begin{align}
    u_1 & = \frac{1}{3}\Bigl( 1+2\cos\frac{\psi}{3}  \Bigr) \label{u1C}\\
    u_2 & = \frac{1}{3}\Bigl( 1+2\cos\frac{\psi-2\pi}{3}  \Bigr)\label{u2C}\\
    u_3 & = \frac{1}{3}\Bigl( 1+2\cos\frac{\psi+2\pi}{3}  \Bigr) \label{u3C}\\
      m & = \frac{1}{2}\left(1-\frac{u_1-(u_2+u_3)/2}{\sqrt{(u_1-u_3)(u_1-u_2)}}\right )\\
      n & = \left((u_1-u_2)(u_1-u_3)\right)^{-\frac{1}{4}} \label{nC}\\
 \chi_i & {\rm ~~such~that~}\  u_i=u_1+\frac{1}{n^2}\hspace{1pt}\frac{1-\cos \chi_i}{1+\cos \chi_i}\ ,
\end{align}
\end{subequations}
%
The type \textbf{{\emph {C}}} differential equation for the orbit is (\refp{orbite}, right):
\beq
u = u_1 + \frac{1}{n^2}\hspace{1pt}
        \frac{ 
	      1-\cn\bigl( F(\chi_i|m) + \frac{\Delta\lambda}{n}|m
	      \bigr) 
	     }{
	      1+\cn\bigl( F(\chi_i|m) + \frac{\Delta\lambda}{n}|m
	      \bigr) 
	      } \label{orbitaC} \ ,
\eq
with $\Delta\lambda = \lambda - \lambda_i$ and $\Delta
\lambda/n~\in~\bigl \lbrace -F(\chi_{BH}\vert m)-F(\chi_i\vert
m)~,~F(\chi_{BH}\vert m)-F(\chi_i\vert m)\bigr\rbrace$, where
$\chi_{BH}=\arccos\frac{1-n^2(1-u_1)}{1+n^2(1-u_1)}$.
%
%
\\
\\
{\bf Critical \emph{C}:}
A special limiting case of solution \refe{orbitaB} for $\mu = 0$ (or \refe{orbitaC} for $\psi=\frac{\pi}{2}$)
($a=\frac{2}{3 \sqrt 3}$) and $u_i>2/3$ ($r_i<3M$) is:
\beq
  u = -\frac{1}{3} +\Cth^2
     \bigl(
           \ArCth\sqrt{u_i+1/3}~+\frac{\Delta\lambda}{2}
     \bigr) \ ,
\label{orbitaKc}
\eq
with $\Delta\lambda = \lambda - \lambda_i$ and $\Delta \lambda ~\in
~\bigl \lbrace 2 \ArCth\sqrt{4/3}-2 \ArCth\sqrt{u_i+1/3}~,~\infty \bigr\rbrace$.

\begin{figure}
\begin{center}
\includegraphics[width=7.7cm]{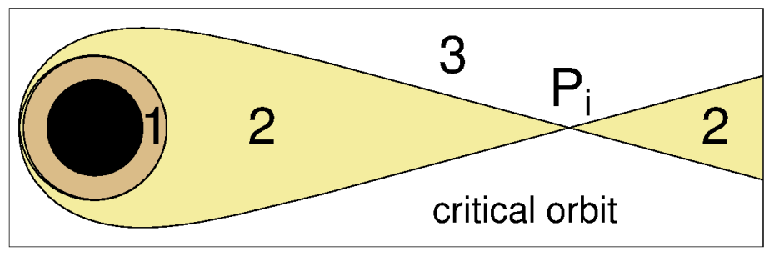}
\esp
\includegraphics[width=5.7cm]{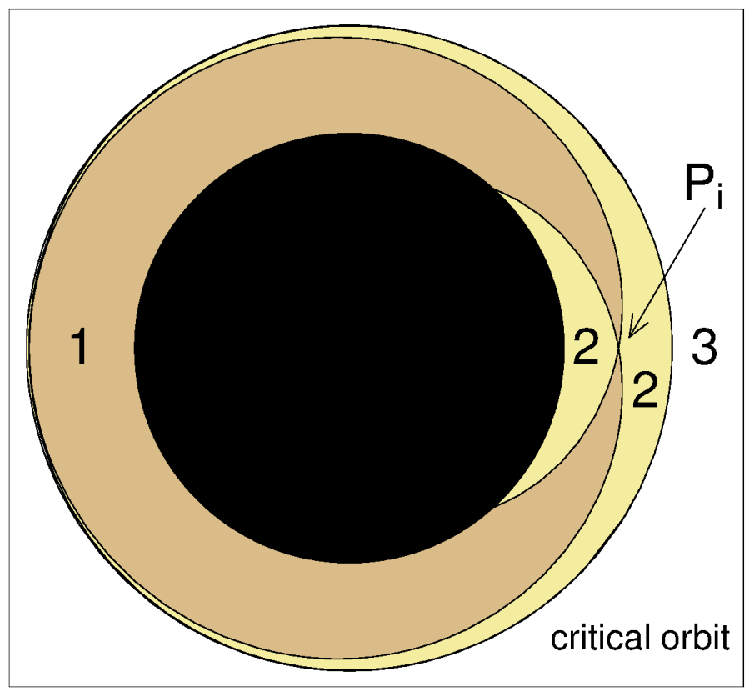}
\end{center}
\caption{For a known initial point $\pp_i$, the two critical orbits
through this point divide the orbital plane into three regions. Left:
initial point at $r>3M$. If the final point
$\pp_f$ is in region 3, then all orbits from $\pp_i$ are of type A. If
$\pp_f$ is in region 1, then all orbits from $\pp_i$ are of type B. If
$\pp_f$ is in region 2, then orbits leading to $\pp_f$ are of type
A if the trajectory winds around the black hole, and of type B if it
goes directly from $\pp_i$ to $\pp_f$. Right: initial point at $r<3M$. If the final point
$\pp_f$ is in region 3, then all orbits from $\pp_i$ are of type B. If
$\pp_f$ is in region 1, then orbits from $\pp_i$ are of type C. If
$\pp_f$ is in region 2, then orbits leading to $\pp_f$ are of type
C if the trajectory winds around the black hole, and of type B if it
goes directly from $\pp_i$ to $\pp_f$.}
\label{podrocja}
\end{figure}

The two constants of motion $\omega$ and $a$ can now be determined
in two steps. First we determine the type of the orbit following
reasoning illustrated in \refp{podrocja} and then we solve the two
equations obtained from the orbit equation at the initial and the final point
\citep{maja, andreja, cadez1, uros}.

We note that the parameter $\omega $ appears in a contrived form as
the starting point of the true anomaly only in the argument of the
Jacobi functions, therefore, it can be eliminated from the two
equations by using the Jacobi elliptic functions addition
theorem. Let $v=F(\chi_i|m)+\Delta\lambda/n$ be the argument of
elliptic functions at the final point of the orbit and $z=F(\chi_i|m)$
be the argument of elliptic functions at the initial point of the
orbit. Then one can use orbit equations  to write:
\\
\noindent
\begin{minipage}[t]{0.35\textwidth}
\begin{subequations}
\begin{align}
\textrm{Type \emph{A}:}\nonumber\\
\cn^2(v|m) & = \frac{u_2-u_f}{u_2-u_3}\label{cnprva}\\
\sn^2(v|m) & = \frac{u_f-u_3}{u_2-u_3}\\
\dn^2(v|m) & = \frac{u_1-u_f}{u_1-u_3}\\
\cn^2(z|m) & = \frac{u_2-u_i}{u_2-u_3}\\
\sn^2(z|m) & = \frac{u_i-u_3}{u_2-u_3}\\
\dn^2(z|m) & = \frac{u_1-u_i}{u_1-u_3}\label{cnzadnja}
\end{align}
\end{subequations}
\end{minipage}
%
%
\begin{minipage}[t]{0.65\textwidth}
\begin{subequations}
\hspace{2.35cm}
\textrm{Types \emph{B} and \emph{C}:}
\begin{align}
\cn(v|m) & = \frac{1-n^2(u_f-u_1)}{1+n^2(u_f-u_1)}\label{cnprvab}\\
\sn^2(v|m) & = \frac{4n^2(u_f-u_1)}{(1+n^2(u_f-u_1))^2}\\
\dn^2(v|m) & =1- \frac{4mn^2(u_f-u_1)}{(1+n^2(u_f-u_1))^2}\\
\cn(z|m) & = \frac{1-n^2(u_i-u_1)}{1+n^2(u_i-u_1)}\label{cnvmesb}\\
\sn^2(z|m) & = \frac{4n^2(u_i-u_1)}{(1+n^2(u_i-u_1))^2}\\
\dn^2(z|m) & =1-
\frac{4mn^2(u_i-u_1)}{(1+n^2(u_i-u_1))^2}\label{cnzadnjab}\ .
\end{align}
\end{subequations}
\end{minipage}
\\
\noindent
Since $v-z=\Delta\lambda_f/n$ one can use the Jacobi
elliptic functions addition theorem to obtain:
\beq
  \cn\left(v-z|m\right)~=~\frac{ 
                     \cn(v|m)\cn(z|m)+ \sn(v|m)\sn(z|m)\dn(v|m)\dn(z|m)
		     }
		     {
		      1-m\hspace{1pt}\sn^2(v|m)\sn^2(z|m)  
		      } \ .
\label{enA6}
\eq
This is a non-linear equation for $a$ (i.e. for $\psi$ or $\mu$
with respect to the type of the orbit). Equations
(\ref{cnprva} -- \ref{cnzadnjab}, except \ref{cnprvab} and
\ref{cnvmesb}) only give squares, so functions $\cn$ and $\sn$ are
determined only up to the sign; $\dn $ is always positive. In order to
determine those signs we plot in \refp{jacobi} the functions $\sn$ and
$\cn$ along orbits together with the interval where the orbit is defined. 
Figure \ref{jacobi} left shows that for type \emph{A} orbits, only the
function $\cn$ changes sign at the periastron and the function $\sn$
is always positive along the orbit. Thus, the sign of $\cn$ is
positive if the orbit did not pass the periaston and negative
otherwise. For orbits of type \emph{B} (\refp{jacobi} middle) $\sn$ is
always positive, while the sign of $\cn$ is determined by
\refe{cnprvab} and \refe{cnvmesb}, thus no sign ambiguity arises. For
orbits of type \emph{C} the sign of $\cn$ is again unambiguous
(\refe{cnprvab} and \refe{cnvmesb}) while the function $\sn$ changes
sign from negative before apastron to positive after apastron (\refp{jacobi} right). 
We conclude that the right hand side of equation \refe{enA6} has two
branches if it contains a sign ambiguity. The cases are:
type \emph{A} ``$\textrm{right}_1$'' and ``$\textrm{right}_2$'' 
and
type \emph{C} ``$\textrm{right}_1$'' and ``$\textrm{right}_3$'', where
\begin{align}
\textrm{right}_1 & =\frac{ 
			   \cn(v|m)\cn(z|m)+ \sn(v|m)\sn(z|m)\dn(v|m)\dn(z|m)
			 }
			 {
			   1-m\sn^2(v|m)\sn^2(z|m)  
			 } \label{desna1}\\
\textrm{right}_2 & =\frac{
                           -\cn(v|m)\cn(z|m)+ \sn(v|m)\sn(z|m)\dn(v|m)\dn(z|m)
			}
			{ 
			   1-m\sn^2(v|m)\sn^2(z|m)  
			} \label{desna2}\\
\textrm{right}_3 & =\frac{
                           \cn(v|m)\cn(z|m)- \sn(v|m)\sn(z|m)\dn(v|m)\dn(z|m)
			}
			{ 
			   1-m\sn^2(v|m)\sn^2(z|m)  
			} \ .\label{desna3}
\end{align}
In the above the sign of $\cn$ or $\sn$ is positive when calculated from a square root. 

\begin{figure}
\begin{center}
\includegraphics[width=5cm]{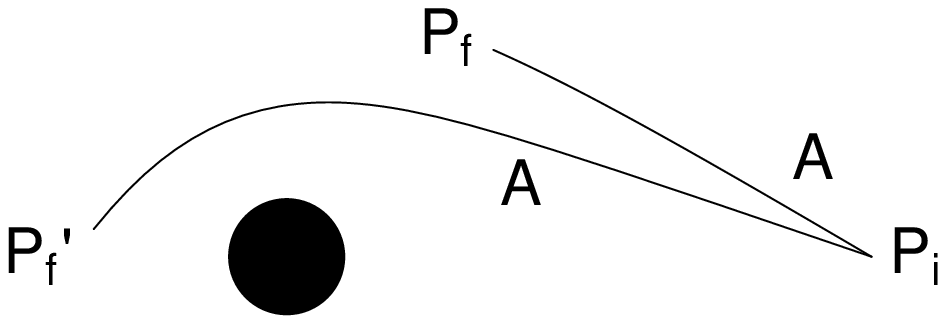}\\
\vspace{0.5cm}
\includegraphics[width=7cm]{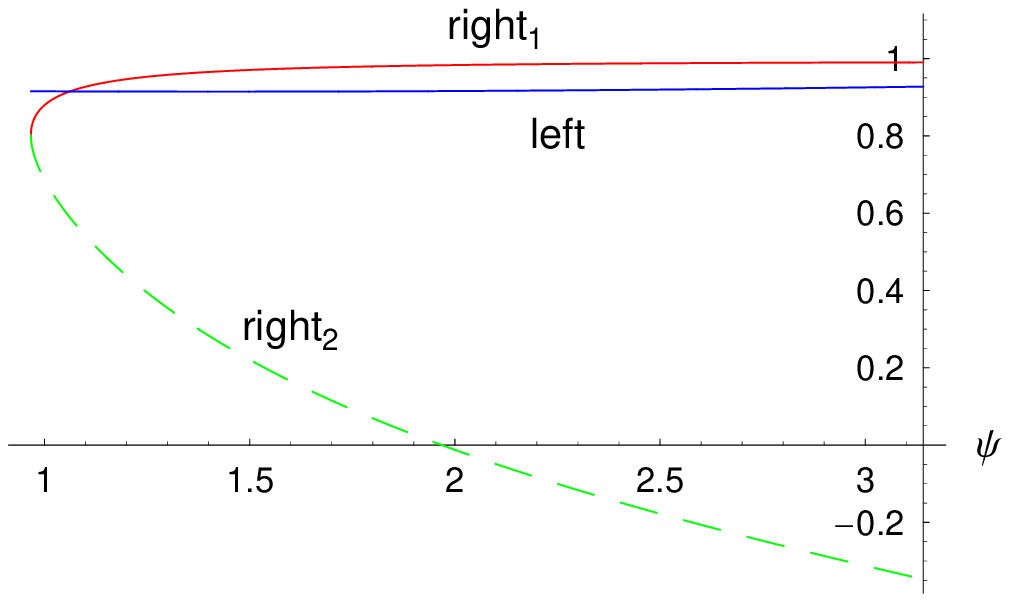}
\esp
\includegraphics[width=7cm]{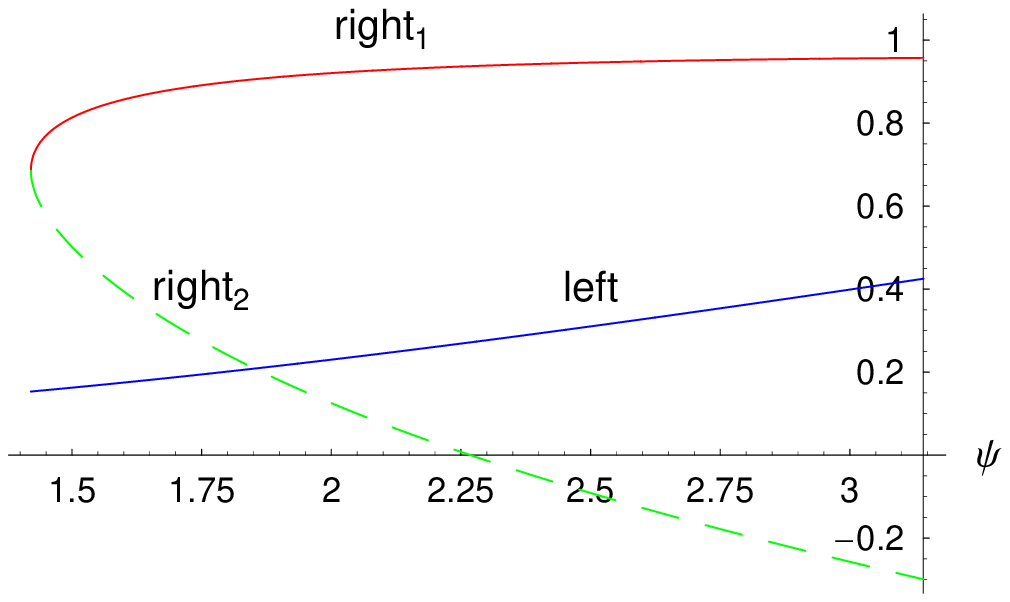}
\end{center}
\caption{Determining the angular momentum parameter for a type \emph{A} orbit.
Left: the orbit from $\pp_i$ to $\pp_f$ does not pass periastron. Right:
the orbit from $\pp_i$ to $\pp_f^\prime$ passes periastron. The
designations $\textrm{right}_1$ and $\textrm{right}_2$ refer to the
branches \refe{desna1} and \refe{desna2} respectively.}
\label{vejeA}
\end{figure}

\begin{figure}
\begin{center}
\includegraphics[width=5cm]{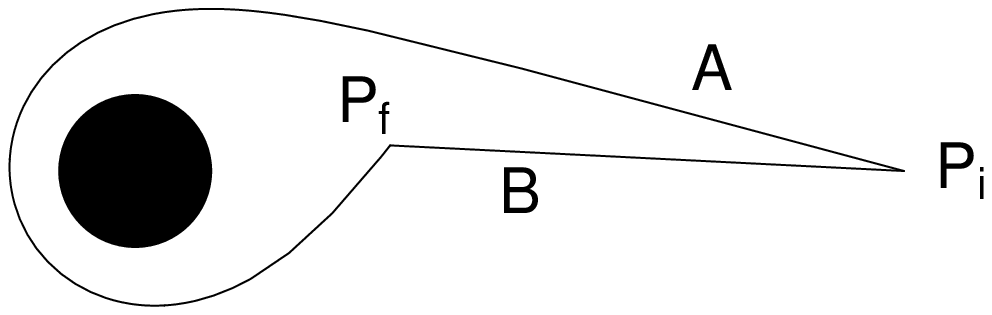}\\
\vspace{0.5cm}
\includegraphics[width=7cm]{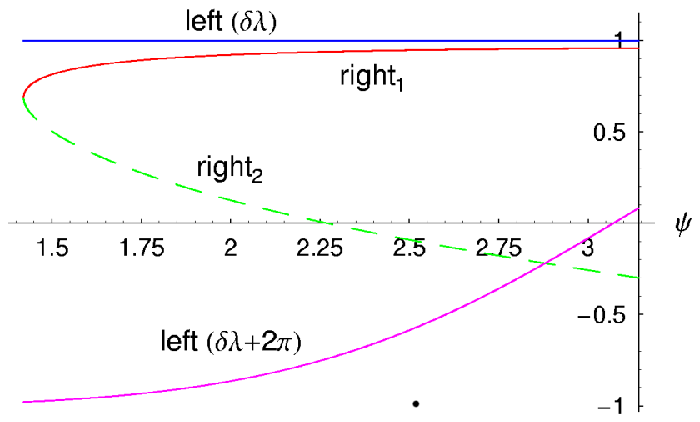}
\esp
\includegraphics[width=7cm]{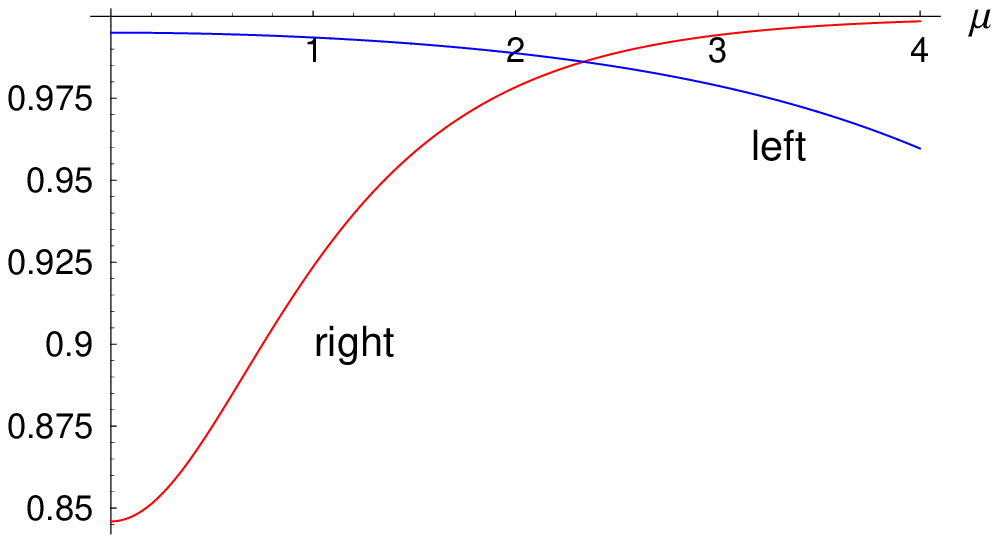}
\end{center}
\caption{Connecting $\pp_i$ and $\pp_f$ with a type \emph{B} orbit and
a type \emph{A} winding orbit (winding number $k=1$); no $k=0$ type
\emph{A} orbit exists in this case. Left: the orbit from $\pp_i$ to
$\pp_f$ of type \emph{A}. Right: the orbit from $\pp_i$ to $\pp_f$ of type \emph{B} .}
\label{vejeAB}
\end{figure}

\begin{figure}
\begin{center}
\includegraphics[width=5cm]{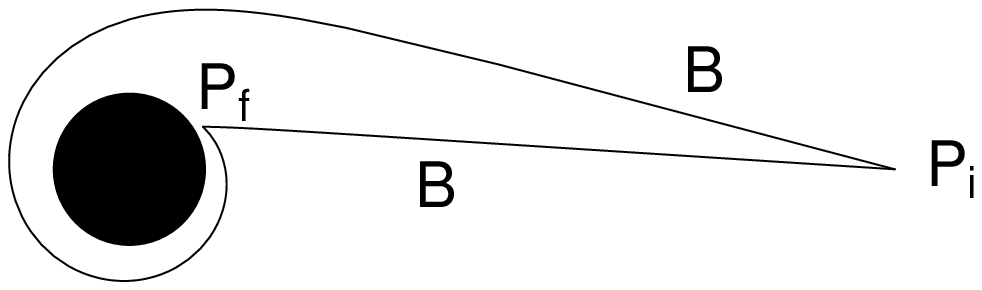}\\
\vspace{0.5cm}
\includegraphics[width=7cm]{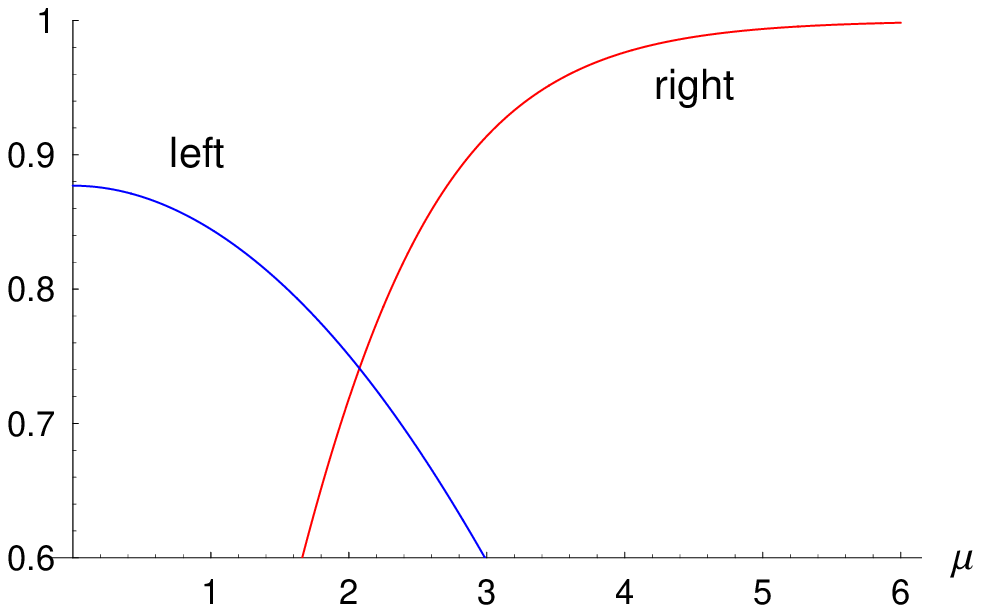}
\esp
\includegraphics[width=7cm]{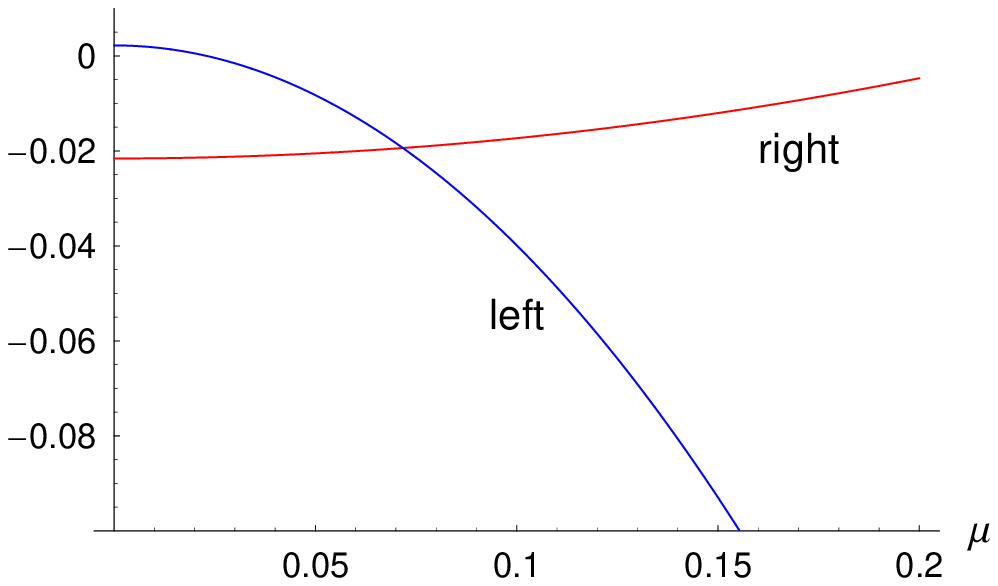}
\end{center}
\caption{
Determining the angular momentum parameter for a type \emph{B} orbit.
Left: a direct orbit from $\pp_i$ to $\pp_f$. Right:
a winding orbit from $\pp_i$ to $\pp_f$ (winding number $k=1$).}
\label{vejeB}
\end{figure}

\begin{figure}
\begin{center}
\includegraphics[width=4cm]{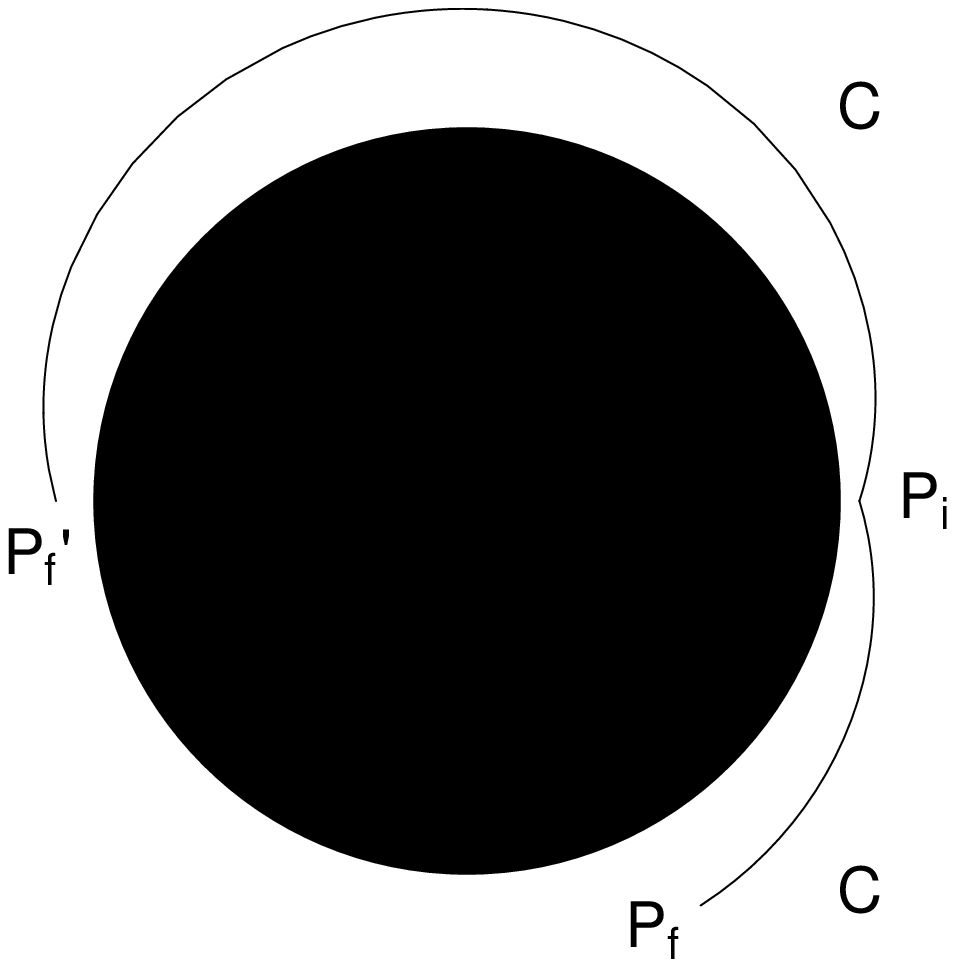}\\
\vspace{0.5cm}
\includegraphics[width=7cm]{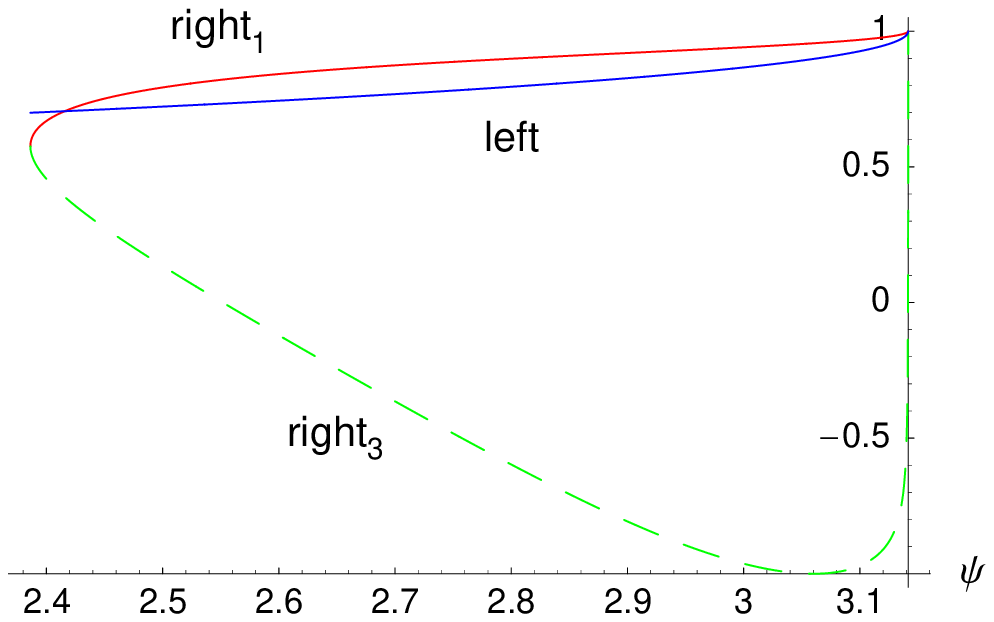}
\includegraphics[width=7cm]{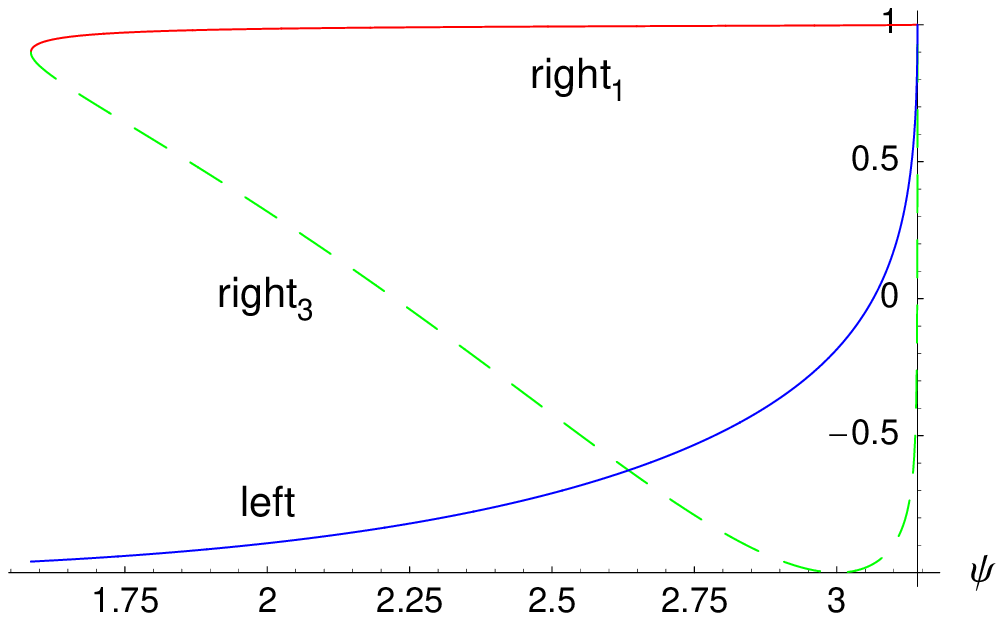}
\end{center}
\caption{Determining the angular momentum parameter for a type \emph{C} orbit.
Left: the orbit from $\pp_i$ to $\pp_f$ does not pass apastron. Right:
the orbit from $\pp_i$ to $\pp_f^\prime$ passes apastron. The
designations $\textrm{right}_1$ and $\textrm{right}_3$ refer to the
branches \refe{desna1} and \refe{desna3} respectively.}
\label{vejeC}
\end{figure}

\begin{figure}
\begin{center}
\includegraphics[width=4cm]{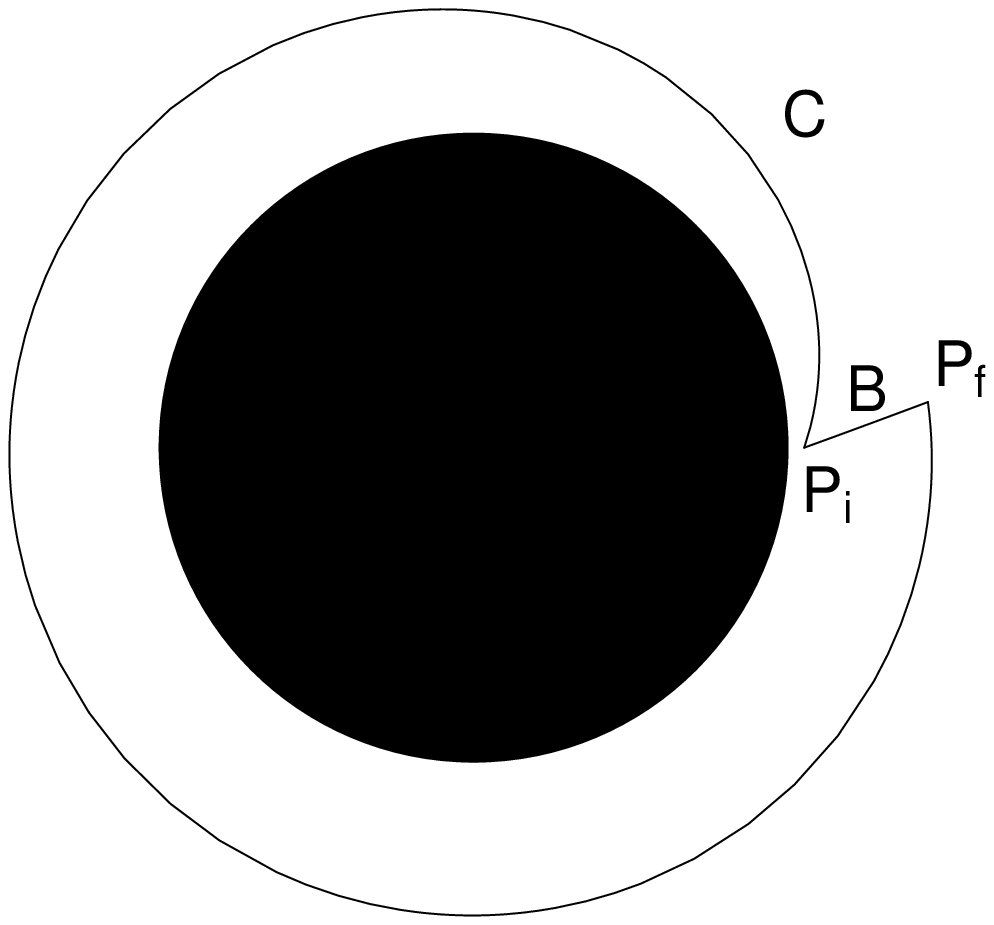}\\
\vspace{0.5cm}
\includegraphics[width=7cm]{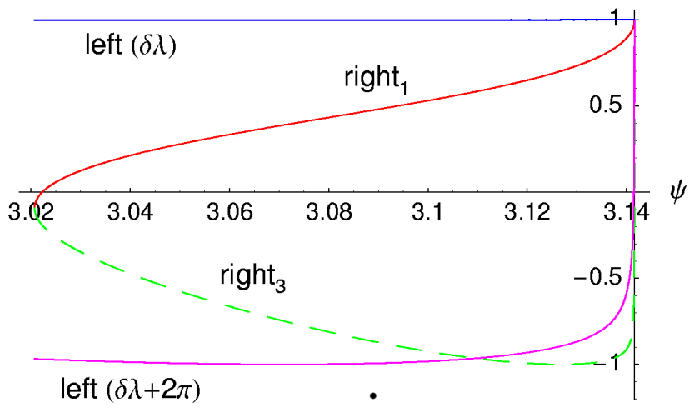}
\includegraphics[width=7cm]{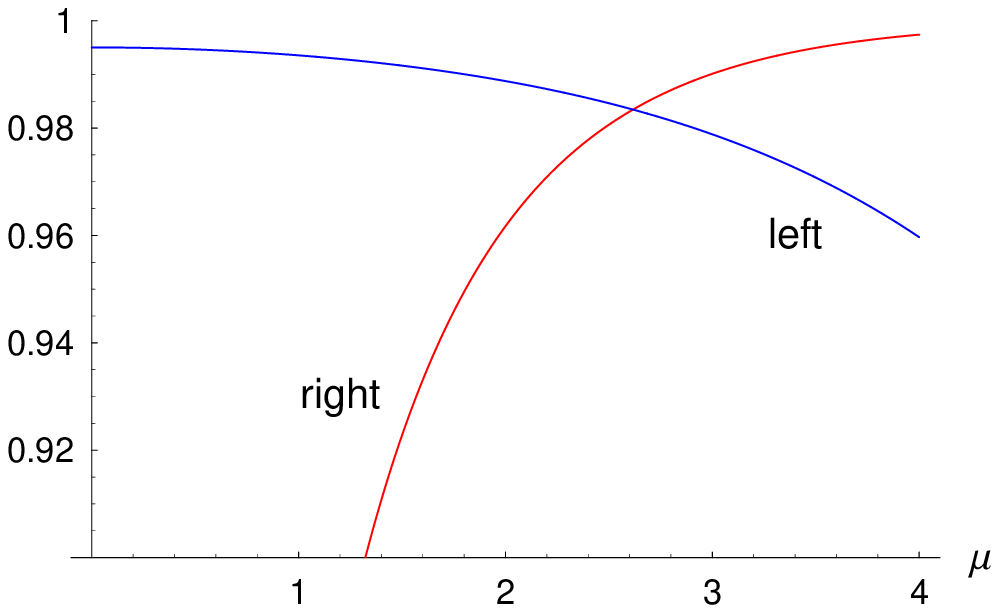}
\end{center}
\caption{
Connecting $\pp_i$ and $\pp_f$ with a type \emph{B} orbit and
a type \emph{C} winding orbit (winding number $k=1$); no $k=0$ type
\emph{C} orbit exists in this case. Left: the orbit from $\pp_i$ to $\pp_f$ of type \emph{C}
. Right: the orbit from $\pp_i$ to $\pp_f$ of type \emph{B} .}
\label{vejeBC}
\end{figure}

Some examples of determining the value of $a$ are shown in figures \ref{vejeA}--\ref{vejeBC}.
They show the left hand side and all the appropriate branches of the right hand side
of \refe{enA6} as a function of $\psi$ or $\mu$
for different cases of $r_i$, $r_f$ and $\Delta \lambda_f$. The
solution of equation \refe{enA6}, which is the point where ``right'' crosses ``left'', is
found numerically by using Brent method
\citep{numericalC}.\footnote{For $\Delta \lambda_f > 2\pi$ there may
be more than one solution to the equation \refe{enA6}; clearly in this
case the orbit is wound about the black hole at $r=3M$, therefore only
the solution with $a$ closest to $\frac{2}{3 \sqrt{3}}$ applies.} With the
now known parameter $a$ it is straightforward to calculate the value
of $\omega$ (equations \ref{orbitaA},\ref{orbitaB},\ref{orbitaC}).

%% file: ms3.tex
The time of flight for a photon, is given by the integral \citep{chandra}:
\beq
  t_{fi} =
     \pm2Ma
     \int_{u_i}^{u_f}
     \frac{d u}{u^2(1-u)\sqrt{a^2-u^2(1-u)}} \ .
\label{cas}
\eq
The indefinite integral in the above formula can be expressed in
analytic form using elliptic integrals. After considerable algebraic
manipulations we obtain the analytic forms for the three 
cases of type \emph{A}, \emph{B} and \emph{C} orbits as follows.
%
\\
\\
{\bf Type \emph{A}:}
Introduce $\chi$ such that (cf \ref{defHiA}) 
\beq
u = u_2-(u_2-u_3)\cos^2\chi \ ,
\eq
to obtain:
\input{ms5.tex}
Here $E$, $F$ and $\Pi$ are the incomplete elliptic integrals of the
first, second and third kind respectively \citep{mathematica}; $u_1$, $u_2$, $u_3$,
$m$ and $n$ are those from (\ref{u1}~--~\ref{nA}), and $n_1$, $n_2$ are:
\begin{align}
 n_1 &= 1-\frac{u_2}{u_3}\\
 n_2 &= \frac{u_2-u_3}{1-u_3}
 \ .
\end{align}
%
\\
\\
{\bf Type \emph{B} and \emph{C}:}
Introduce $\chi$ such that (cf \ref{hib}) 
\beq
u = u_1+\frac{1}{n^2}\hspace{1pt}\frac{1-\cos \chi}{1+\cos \chi} \ ,
\eq
to obtain:
\input{ms6.tex}
where
\begin{align}
\alpha_1 & = \frac{n^3}{n^2 u_1+1} \label{alfa1}
\esp
\alpha_2 = \frac{n^5}{(n^2 u_1+1)^2}
\esp
\alpha_3 = \frac{n^3}{n^2(1-u_1)-1}\\
k_1 & = \frac{1 - u_1 n^2}{1 + u_1 n^2}
\esp
k_2 = \frac{1+(1-u_1)n^2}{1-(1-u_1)n^2}\\
n_1 & = \frac{k_1^2}{k_1^2-1}
\esp
n_2 = \frac{k_2^2}{k_2^2-1}\\
x_1 & = \frac{
              \sqrt{n_1-m}\sin\chi+\sqrt{1-m\sin^2 \chi}
              }{
	      \sqrt{n_1-m}\sin\chi-\sqrt{1-m\sin^2 \chi}
              }\\
x_2 & = \frac{
              \sqrt{n_2-m}\sin\chi+\sqrt{1-m\sin^2 \chi}
              }{
	      \sqrt{n_2-m}\sin\chi-\sqrt{1-m\sin^2 \chi}
              } \label{x2} \ .
\end{align}
For type \emph{B} orbits the parameters $u_1$, $m$ and $n$ are those
from (\ref{u1B}~--~\ref{nB}), while for type \emph{C} orbits these
parameters are from (\ref{u1C}~--~\ref{nC}).

\begin{figure}
\begin{center}
\includegraphics[width=10cm]{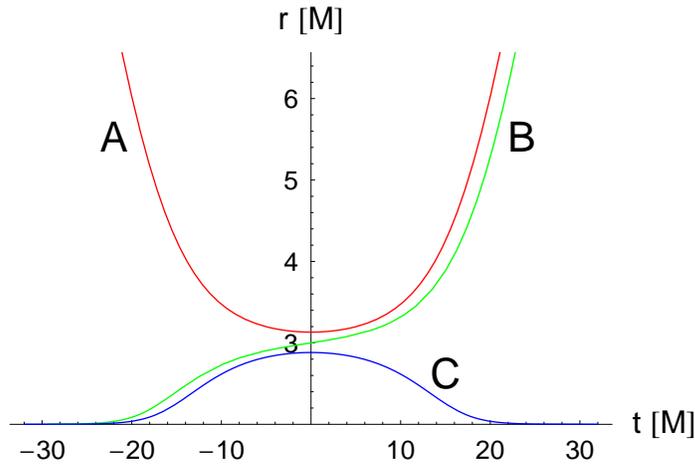}
\end{center}
\caption{Graph $r=r(t)$ in units of $M$ for all three types of
orbits. The type \emph{A} orbit starts at infinity at 
$t\rightarrow -\infty$, passes the periastron at  $t=0$ and
continues to infinity as $t\rightarrow \infty$. The type \emph{B}
orbit starts at the event horizon at $t\rightarrow -\infty$,
crosses the critical $r=3M$ at $t=0$ and continues to infinity
as $t\rightarrow \infty$. The type \emph{C} orbit starts at the event
horizon at $t\rightarrow -\infty$, passes the apastron at 
$t=M$ and again approaches the event horizon as $t\rightarrow
\infty$. In the above example $a=\frac{2}{3\sqrt{3}}-10^{-3}$ for type \emph{A} and \emph{C}
orbits and $a=\frac{2}{3\sqrt{3}}+5 \cdot 10^{-4}$ for type \emph{B} orbit.}
\label{casi}
\end{figure}

In figure \ref{casi} we show examples of $r(t)$ for all three types of orbits.
We compared the effectivness of the analytic method and the pure
numerical method to calculate the times of flight. It was found that
the algorithm based on the analytic method and implemented with the
Carlson's algorithm \citep{Carlson} is always more accurate and 3 to 6 times faster than \mbox{fourth-order}
\mbox{Runge-Kutta} integration with adaptive stepsize
control.\footnote{We note that the algorithm \emph{ellpi} of Numerical
Recipes \citep{numericalC} does not give the same results as
Mathematica for $\Pi$ integrals. We rewrote the function \emph{rj}
according to Carlson's original paper \citep{Carlson}, and obtained
identical results with Mathematica.} We also note
\citep{uros} that it is possible to calculate the time of flight
numerically by expanding the integrand \refe{cas} into piecewise
rapidly convergent series of analytically integrable functions. Such
series easily give results accurate to $10^{-8}M$ for subcritical
orbits and are some 6 times faster than the analytic method. The most
effective method to calculate the time of flight would thus be a
combination of the analytic method for the close to critical orbits
and the series solution for the rest.

%% file: ms5.tex
\begin{equation}
\begin{split}
t(\chi) & = \frac{2an}{u_3^2}\Biggl[
\left(
1+u_3+\frac{  n_1^2-m  }{  2(m-n_1)(n_1-1)  }
\right)\Pi(n_1;\chi|m) +
\frac{u_3^2}{1-u_3}\Pi(n_2;\chi|m) \\
  &\quad +
\frac{n_1/2}{(m-n_1)(n_1-1)}
\left(
E(\chi|m)-\Bigl(1-\frac{m}{n_1}\Bigr)F(\chi|m)
-
\frac{
n_1\sin 2\chi\sqrt{1-m\sin^2 \chi}
}{
2(1-n_1\sin^2 \chi)
}
\right)
\Biggr] \ ,
\label{resiteva}
\end{split}
\end{equation}

%% file: ms6.tex
\begin{align}
\begin{split}
t(\chi) & = 2a\Biggl[
\frac{\alpha_2(n_1-1)}{n_1^2+m-2mn_1}
   \left(
     \frac{
       n_1\sin\chi(k_1\cos\chi+1)\sqrt{1-m\sin^2\chi}
     }{
       k_1(1-n_1\sin^2\chi)
     }
     -E(\chi|m)
   \right)\\
 &\quad +\alpha_3(1-n_2)(1+\frac{1}{k_2})
    \left(
      \Pi(n_2;\chi|m)+\frac{k_2}{2\sqrt{n_2-m}}\ln|x_2|
    \right)\\
 &\quad +\alpha_1(1-n_1)(1+\frac{1}{k_1})
    \left(
      \Pi(n_1;\chi|m)+\frac{k_1}{2\sqrt{n_1-m}}\ln|x_1|
    \right)\\
 &\quad +F(\chi|m)
    \left(
       \frac{\alpha_2}{k_1^2}(1+(1+k_1)^2(n_1-1))-\frac{\alpha_1}{k_1}-\frac{\alpha_3}{k_2}
    \right)
\Biggr]
\ ,
\label{resitevb}
\end{split}
\end{align}

%% file: ms4.tex
In this work we presented the complete solution of the ray-tracing problem in the Schwarzschild space-time. All the algorithms presented here have been tested for accuracy and for speed of execution and were found to be more accurate and considerably faster than commonly used direct integration methods.
We hope that algorithms presented here will be used as a useful tool in solving more complex ray tracing problems that will elucidate the physics governing complicated transient phenomena in the vicinity of black holes. We would like to remind the community that a similar ray-tracing problem in the Kerr space-time still remains to be solved.